\documentclass{article}
\pdfoutput=1
\usepackage{jheppub}
\usepackage{amsmath,hyperref,amssymb}
\usepackage{graphicx}
\usepackage{dcolumn}

\usepackage{bm}
\usepackage{verbatim}
\usepackage{tabularx}
\usepackage{slashed}
\usepackage{float}

\def\be{\begin{equation}}
\def\ee{\end{equation}}
\def\bea{\begin{eqnarray}}
\def\eea{\end{eqnarray}}

\newcommand{\pq}{$U(1)_{PQ}$ }
\definecolor{seagreen}{rgb}{0.180392,0.545098,0.341176}

\newcommand*\br[1]{\overline{#1}} 
 
\newcommand*\gm{\gamma^\mu}
\newcommand*\gf{\gamma^5}

\begin{document}

\title{Light axions with large hadronic couplings}
\author{Gustavo Marques-Tavares,}
\author{Mae Teo}
\affiliation{Stanford Institute for Theoretical Physics, Stanford
University, Stanford, CA 94305, USA}

\abstract{ 
  
We present models in which axions have large couplings to hadrons while remaining naturally light.
By assigning \pq charges to Standard Model quarks such that \pq is not anomalous under QCD, the derivative couplings naturally arise while no potential is generated for the axion upon QCD confinement. We present simple models that implement this idea both for an axion-like particle and for the QCD axion. We show that both models lead to sizable flavor violations that could be probed by future experiments. Our construction shows that the axion coupling to hadrons can be essentially independent from its mass and thus motivates experimental searches in a wide range of axion parameter space.
}

\maketitle

\section{Introduction}

Axions are one of the most well-motivated candidates for physics beyond the Standard Model. They arise generically in string theory~\cite{Svrcek:2006yi,Arvanitaki:2009fg}, and are also excellent dark matter candidates~\cite{Abbott:1982af,Dine:1982ah,Preskill:1982cy,Arvanitaki:2009fg,Arias:2012az}. Axions can also provide an elegant solution to the strong CP problem by promoting the QCD $\theta$ parameter to a dynamical degree of freedom which is dynamically driven to zero~\cite{PhysRevLett.38.1440,PhysRevLett.40.223,PhysRevLett.40.279}. In the remainder of this paper we will refer to the axion that solves the strong CP problem as the \textit{QCD axion}, and use \textit{axions} for all other cases (also called axion-like particles or ALPs in the literature).

There is a growing number of experimental proposals to search for axions covering a wide range of parameter space. Some of the proposals search for the axion via its coupling to photons, and can be divided by proposals that require the axion to be a sizable fraction of dark matter, e.g. ADMX \cite{RYBKA201414}, HAYSTAC \cite{PhysRevLett.118.061302}, ABRACADABRA \cite{PhysRevLett.117.141801}, DM radio \cite{Silva-Feaver:2016qhh}; and experiments that look for axions produced in the lab or by astrophysical objects, e.g. CAST \cite{Anastassopoulos:2017ftl}, ALPS-II \cite{Bahre:2013ywa}, IAXO \cite{1475-7516-2011-06-013} (see Figure \ref{fig:gaGG} for a more complete list of experiments). Another class of proposals will search for the axion via its couplings to hadrons. They can also be classified by either requiring the axion to be a sizable fraction of dark matter, e.g. CASPEr-Wind, CASPEr-Electric~\cite{Graham:2013gfa,JacksonKimball:2017elr}, nEDM \cite{Abel:2017rtm}, and recent proposals that make use of torsion pendulums and atomic magnetometry \cite{Graham:2017ivz};\footnote{The axion can have two separate hadronic couplings: one through the $a G \tilde G$ operator, which induces a neutron electric dipole moment in the background of an axion field, and a derivative coupling to nucleons. CASPEr-Wind and the experiments with torsion pendulums and magnetometry are sensitive to the former, while CASPEr-Electric and nEDM probe the latter.
}  or by searching for axions sourced in the lab, e.g. ARIADNE \cite{Arvanitaki:2014dfa}.

Given the exciting experimental program searching for axions, it is important to explore which models are theoretically motivated. While it is straightforward to construct models in which the axion-photon coupling is effectively a free parameter, the coupling to hadrons generally implies additional properties of the axion. In generic models of axions that couple to hadrons, a non-perturbative QCD potential is generated for the axion because hadrons are composite states of fields charged under QCD. In such cases, for a given coupling to hadrons, $g_{aNN}$, there is an implied (approximate) minimal mass for the axion given by this non-perturbative potential. The overall relation between $g_{aNN}$ and $m_a$ is model dependent and can change by order one factors, but is expected to be parametrically given by $m_a \sim g_{aNN} m_\pi f_\pi$. One way to evade this relation and have a parametrically smaller mass is to have an extra contribution to the potential which is tuned to partially cancel the potential generated by QCD. This possibility has a number of interesting phenomenological consequences which were recently discussed in Ref.~\cite{Hook:2017psm} (see also Ref. \cite{Hook:2018jle} for a natural mechanism to achieve such a cancellation for the mass).

In this work we explore a class of models in which the hadronic couplings of the axion do not lead to an axion mass. In these models, some Standard Model quarks are necessarily charged under the $U(1)$ symmetry associated with the axion (which we will refer to as \pq), but with charge assignments such that this symmetry is not anomalous under QCD.\footnote{
In the last part of this work we will also consider cases in which \pq is anomalous under QCD but only due to states beyond the Standard Model.
} In such scenarios, once the \pq is broken there are derivative couplings between quark currents and the axion, which induce axion-hadron couplings. However, since \pq is not anomalous under QCD, the axion shift symmetry is preserved and no potential is generated.

Because existing experimental constraints require the decay constant of the axion to be $f \gg v$, where $v$ is the Higgs vacuum expectation value (VEV), the axion cannot be associated with a degree of freedom of a two Higgs doublet model as in the original PQWW QCD axion model \cite{PhysRevLett.38.1440,PhysRevLett.40.223,PhysRevLett.40.279}. It follows that at energies above $f$, the Yukawa couplings of any quark field charged under \pq must come from a higher dimensional operator. Given the large value of the top quark Yukawa, we will assume that the 3rd generation quarks are neutral under \pq. This means that the coupling to the axion breaks the flavor symmetries of the quark sector and can mediate flavor changing neutral current (FCNC) interactions. Such interactions are strongly constrained, especially in the first two generations. In order to have the largest possible hadronic couplings allowed by FCNC constraints, we will focus on a model in which the \pq charge assignments respect a $U(2)\times U(2)$ flavor symmetry of the first two generations. This will lead to interesting connections between the size of the axion-hadron couplings and the expected meson decay widths from FCNCs (see also \cite{Ema:2016ops,Calibbi:2016hwq,Alves:2017avw,DiLuzio:2017ogq} for other recently studied axion models with flavor non-diagonal couplings).

In the next section, we present the model and work out the relevant couplings between the axion and Standard Model particles. After that, we show the allowed parameter space for such models and the projected sensitivity of proposed experiments. In the last section, we present a model of the QCD axion with hadronic couplings parametrically larger than what would be expected in generic models.  

\section{Model}

The motivation of the model is to generate shift-symmetric axion couplings to quarks without simultaneously generating the $a G \tilde G$ coupling to gluons which breaks the axion shift symmetry. This can be achieved by assigning \pq charges to the SM quarks in a way that does not make \pq anomalous under QCD. In this work we focus on a scenario where the \pq charge assignment respects a $U(2)\times U(2)$ flavor symmetry of the Standard Model, with only the first two generations charged under \pq. The \pq charge assignment of the first two generations is given in Table~\ref{tab:charges}. One can easily check that given these charges, \pq is not anomalous under QCD.

\begin{table}
\begin{center}
\begin{tabular}[hbt]{c | c}
	 &	\pq  \\ \hline
	 $u^c_{i=1,2}$ & $+1$ \\
	 $d^c_{i=1,2}$ & $-1$ \\
	 $Q_{i=1,2}$ & $0$
\end{tabular}
\caption{\pq charge assignments for the first two generations, where we have taken all fields to be left-handed spinors. The third generation quarks are neutral under \pq.}
\label{tab:charges}
\end{center}
\end{table}

In addition to the SM fields the model has a scalar field $\phi$ which has charge $-1$ under $U(1)_{PQ}$. We assume that this field gets a VEV $f$ which breaks the PQ symmetry and that the axion is the corresponding Nambu-Goldstone boson. The Yukawa interactions involving the first two generation singlets are forbidden by \pq and therefore must arise from dimension five operators involving $\phi$. In this case the quark interactions with the Higgs are given by
\begin{equation}
	\mathcal{L} \supset \sum_{i=1}^{3} \sum_{j=1}^2 \left( \lambda^u_{ij} \frac{\phi}{\Lambda} Q_i H u^c_j + \lambda^d_{ij} \frac{\phi^\dagger}{\Lambda} Q_i \tilde H d^c_j \right) +
	 \sum_{i=1}^3 \left( y^u_{i 3} Q_i H u^c_3 + y^d_{i 3} Q_i \tilde H d^c_3 \right) + h.c.,
	 \label{eq:lagrangian-symmetry}
\end{equation}
where $\Lambda$ is the cutoff scale for this effective theory and should be larger than $f$, the VEV of the scalar field $\phi$.\footnote{
For a possible UV completion of this dimension 5 operator see e.g. Ref.~\cite{DiLuzio:2017ogq}.
}

In the broken phase of \pq, we can replace $\phi \rightarrow  \tfrac{f}{\sqrt{2}} e^{-i a/f} $. At energies below $f$ the quark Yukawa sector is given by
\begin{equation}
	\mathcal{L} \supset \sum_{i=1}^{3} \sum_{j=1}^2 \left( y^u_{ij} e^{-i a/f} Q_i H u^c_j + y^d_{ij} e^{i a/f} Q_i \tilde H d^c_j \right) +
	 \sum_{i=1}^3 \left( y^u_{i 3} Q_i H u^c_3 + y^d_{i 3} Q_i \tilde H d^c_3 \right) + h.c.  ,
\end{equation}
where $y^{u/d}_{ij}$ are the quark Yukawa couplings in the Standard Model. Notice that if we assume a hierarchy between $f$ and $\Lambda$, as one would expect if the Lagrangian in Eq.~\ref{eq:lagrangian-symmetry} is valid over a significant energy range, the entries $y_{ij}$, for $j=1,2$, in the Yukawa matrices are suppressed by $f/\Lambda \equiv \epsilon$. This structure of the Yukawa matrices can partially explain the flavor hierarchy between the third generation and the first two, if one assumes that the $\lambda_{ij}$ and $y_{i3}$ in Eq.~\ref{eq:lagrangian-symmetry} are of the same order. Motivated by this, we take the small parameter $\epsilon$ to be roughly in the range $m_s/m_b \lesssim \epsilon \lesssim 1/4\pi$. 

After electroweak symmetry breaking, the axion couples to the quarks as phases in some entries of the quark mass matrix. The field redefinition $u^c_i \rightarrow e^{i a/f} u^c_i$ and $d^c_i \rightarrow e^{-i a/f} d^c_i$, for $i=1,2$, removes the quark-axion couplings from the quark mass matrix and generates the following couplings
\begin{equation} \label{couplings1}
	\mathcal{L} \supset +\frac{\partial_\mu a}{f} \sum_{i=1}^2 \left( u^{c \dagger}_i \sigma^\mu u^c_i - d^{c \dagger}_i \sigma^\mu d^c_i \right) + \frac{\alpha}{2 \pi} \frac{a}{f} F \tilde F \, ,
\end{equation}
where $F$ is the photon field strength. These couplings break the flavor symmetry and when written in the mass basis they give rise to flavor changing interactions suppressed by $\epsilon$:
\begin{equation}
\begin{aligned}
	\mathcal{L} \supset & +\frac{\partial_\mu a}{f} \sum_{i=1}^2 \left[ \,u^{c \dagger}_i \sigma^\mu u^c_i - d^{c \dagger}_i \sigma^\mu d^c_i \right] - \epsilon\frac{\partial_\mu a}{f} \sum_{i=1,2}\left( c^u_{i3} u^{c \dagger}_3 \sigma^\mu u^c_i - c^d_{i3} d^{c \dagger}_3 \sigma^\mu d^c_i  + h.c. \right) \\
	& - \epsilon^2 \frac{\partial_\mu a}{f} \sum_{i,j=1}^2 \left(c^u_{ij} u^{c \dagger}_i \sigma^\mu u^c_j - c^d_{ij} d^{c \dagger}_i \sigma^\mu d^c_j  + h.c. \right) \, ,
\end{aligned}
\label{eq:couplings-mass-basis}
\end{equation}
where we expect $|c_{ij}| \sim \mathcal O(1)$. 

The strongest constraint on such flavor changing neutral current interactions comes from meson decays, in particular $K^\pm \rightarrow \pi^\pm a$, which in this model has a decay width given by
\begin{equation}
	\Gamma(K^\pm \rightarrow \pi^\pm a) = |c_{12}^d|^4 \epsilon^4 \frac{1}{64\pi} \frac{m_K^3}{f^2} \left(1- \frac{m_\pi^2}{m_K^2} \right)^3 \,.
\end{equation}
The experimental bounds on this decay is $\text{Br}(K^\pm \rightarrow \pi^\pm \text{ inv}) < 0.73 \times 10^{-10}$~\cite{Adler:2008zza}, which implies 
\begin{equation}
f \gtrsim 1 \times 10^{8} \text{ GeV} \times \left(\frac{\epsilon}{0.02}\right)^2 \, ,
\end{equation} 
if we take $|c_{12}^d|=1$. The NA62 collaboration is also sensitive to this decay~\cite{na62} and is expected to improve this bound.
Another interesting constraint comes from the decay $B^\pm \rightarrow K^\pm a$, which currently sets a bound at $f > 6\times 10^6  \text{ GeV} \times \left( \frac{\epsilon}{0.02} \right)$. The Belle II experiment \cite{Abe:2010gxa} is expected to improve the bound on this decay by one or two orders of magnitude, which would make the sensitivity of $B^\pm$ decays to axions comparable to that of Kaon decays.

Because the interactions in Eq.~\ref{eq:lagrangian-symmetry} do not generate a mass for the axion, there must be another source of \pq breaking in order to give the axion a mass. For an axion that doesn't solve the strong CP problem, the mass can be generated by hidden sector dynamics that breaks the continuous shift symmetry associated with the axion. For example, a gauge group under which \pq is anomalous with confinement scale $\Lambda_h$ would generate a potential 
\begin{equation}
V  \approx \Lambda^4_h \cos(a/f) \,,
\label{eq:alp-potential}
\end{equation}
giving the axion a mass $m \approx \Lambda^2_h/f$. We explore a mechanism for generating a mass for the QCD axion that is independent of the size of the derivative coupling to quarks in Section~\ref{section:clockwork}.

\begin{figure}[t]
  \centering
    \includegraphics[width=\linewidth]{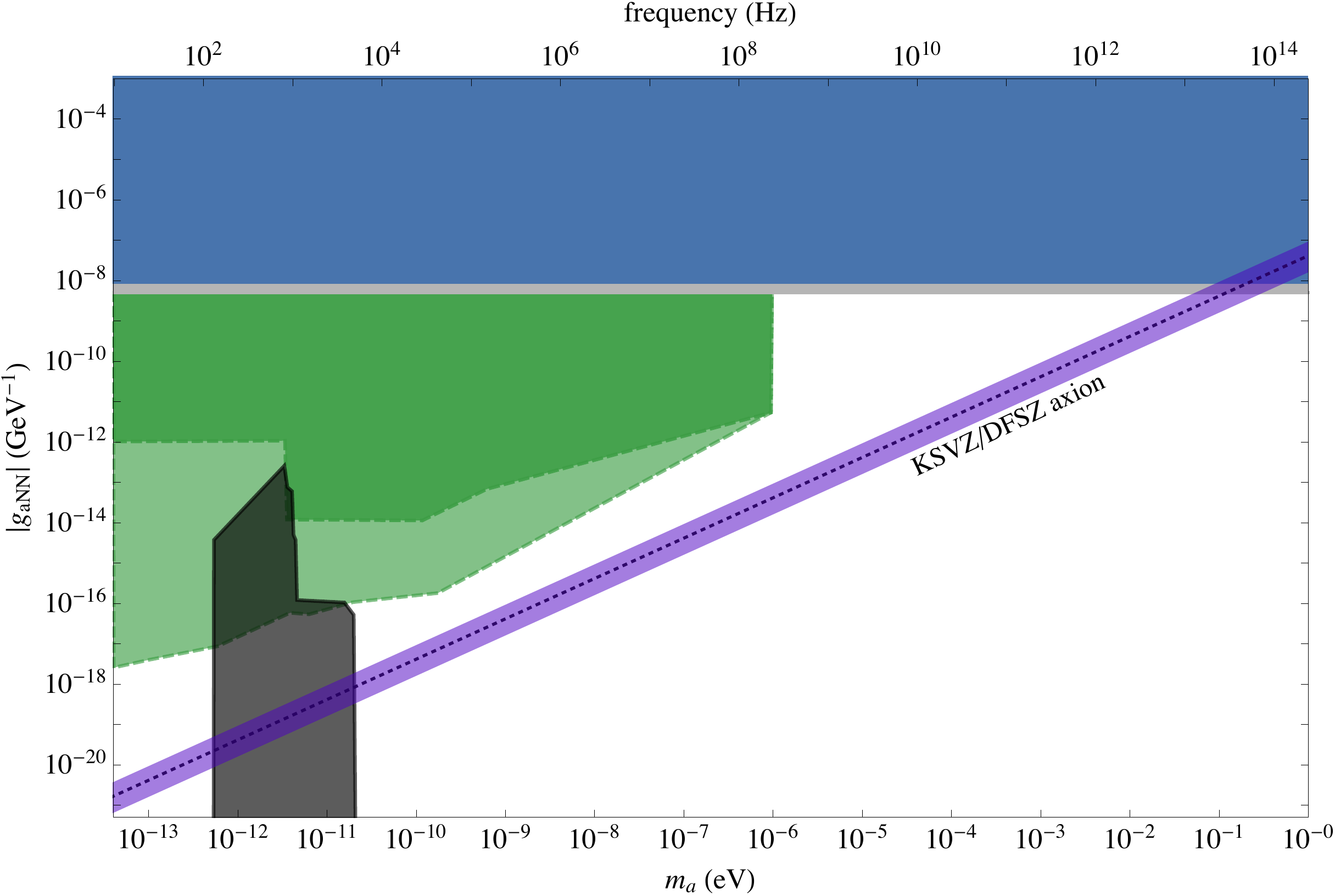}\\
   \caption{Current and future bounds on the axion-nucleon coupling. Regions bounded by solid lines are existing constraints from the branching ratio of $K\rightarrow \pi a$  \textit{(blue)} (with $|c_{12}|=1$ and $\epsilon = 0.02$), SN1987A \cite{Chang:2018rso} \textit{(grey)} , and black hole superradiance \cite{Arvanitaki:2014wva} \textit{(black)} . The darker green region will be probed by CASPEr-Wind \cite{JacksonKimball:2017elr}, and the lighter green region shows its ultimate theoretical reach \cite{Graham:2013gfa}. The KSVZ axion line is shown in black (dotted), while the DFSZ axion band is shown in purple. The axion-nucleon coupling has the standard relation for these axions: $g_\text{aNN} \sim 1/f_a \sim m_a/\Lambda_\text{QCD}^2$. (To be concrete, we have plotted the axion-proton coupling given in \cite{diCortona:2015ldu}.) In our model, $g_\text{aNN} \sim 1/f$, where $f$ can be an arbitrary value, allowing $g_\text{aNN}$ to be parametrically larger. }
  \label{fig:gaNN}
\end{figure}

The axion coupling to quarks in Eq.~\ref{eq:couplings-mass-basis} leads to its coupling to nucleons, defined by $\mathcal{L} \supset g_{\text{aNN}} (\partial_\mu a) \br{N} \gm \gf N$ with~\cite{Kaplan:1985dv,diCortona:2015ldu}
\begin{align}
g_{\text{aNN}} =\pm \frac{1.27}{2f} \sim \pm \frac{1}{f} \,.
\end{align}
The coupling is $+(-)$ for protons (neutrons), and here we have neglected $\mathcal{O}(\epsilon^2)$ effects. Interestingly, the axion couples to protons and neutrons with opposite signs, from the derivative coupling to quarks being proportional to $\tau_3$ of isospin. Since $g_{\text{aNN}}$ is independent of the mass of the axion, it can be significantly larger than the corresponding $g_{\text{aNN}}$ for a typical KSVZ/DFSZ axion. This increases considerably the range of theoretically motivated parameter space that should be experimentally tested.  Current and future experimental bounds on $g_{\text{aNN}}$ are shown in Figure \ref{fig:gaNN}. 


%
\begin{figure}[t]
  \centering
    \includegraphics[width=\linewidth]{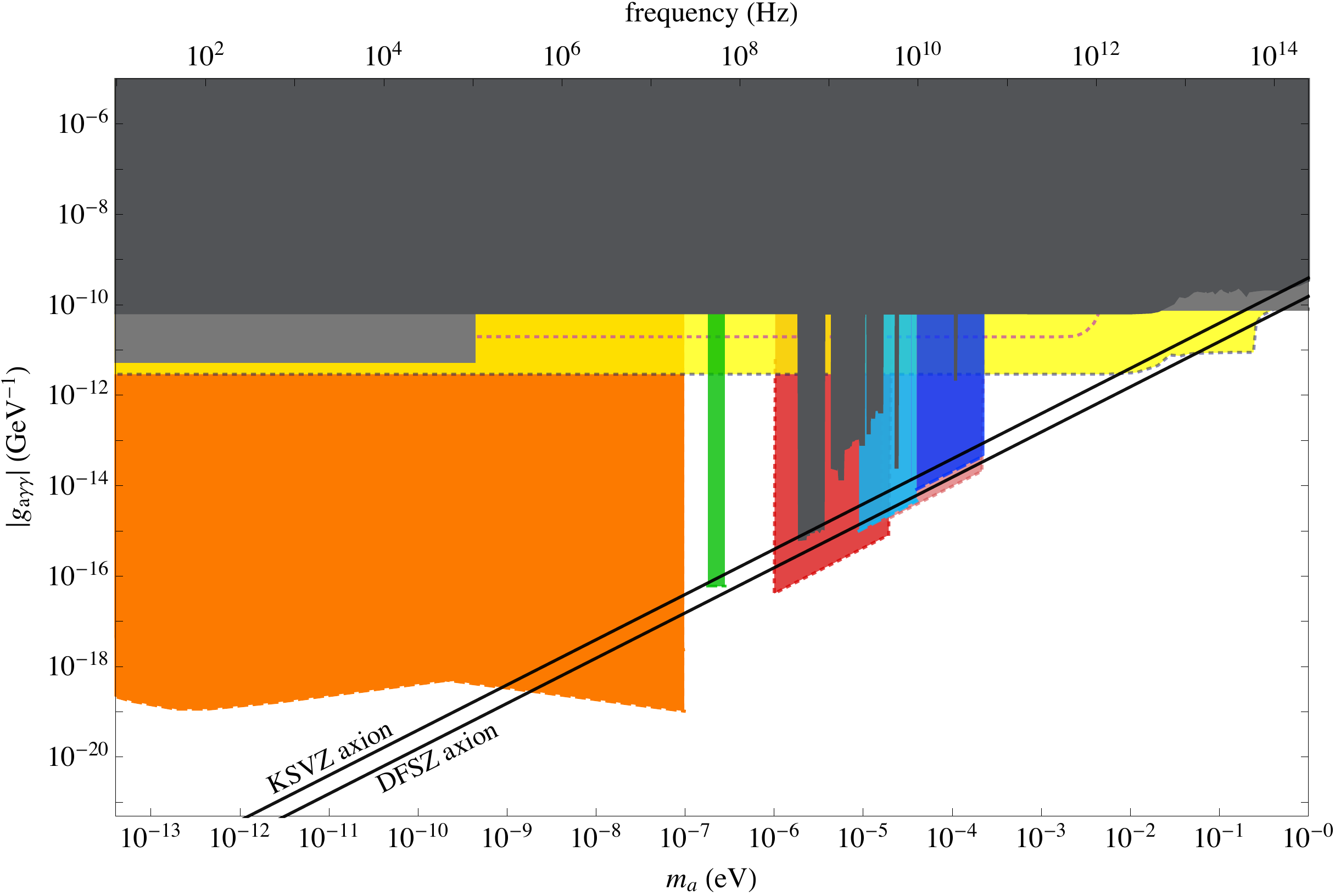}\\
   \caption{Existing and future constraints on the axion-photon coupling. The lighter grey regions are existing constraints from horizontal branch stars \cite{PhysRevLett.113.191302} and SN1987A \cite{1475-7516-2015-02-006}, while the darker grey regions are those from CAST \cite{Anastassopoulos:2017ftl}, ADMX  \cite{Asztalos:2003px,PhysRevLett.104.041301}, RBF and UF \cite{PhysRevLett.59.839,PhysRevD.40.3153,Hagmann:1990tj}, HAYSTAC \cite{PhysRevLett.118.061302}, and ORGAN \cite{McAllister:2017lkb}.
   Future constraints are expected from proposed and ongoing experiments:   IAXO \cite{1475-7516-2011-06-013} \textit{(yellow)}, ALPS-IIc \cite{Bahre:2013ywa} \textit{(pink dotted line)}, ADMX \cite{RYBKA201414} \textit{(red)}, ADMX-HF \cite{Shokair:2014rna} \textit{(pink)}, CULTASK \cite{Chung:2016ysi} \textit{(light blue)}, RADES (prototype) \cite{Melcon:2018dba} \textit{(purple)}, MADMAX \cite{PhysRevLett.118.091801} \textit{(dark blue)}, KLASH \cite{Alesini:2017ifp} \textit{(green)}, and ABRACADABRA \cite{PhysRevLett.117.141801} \textit{(orange)}.
The black lines show the expected axion-photon coupling for the KSVZ and DFSZ axion, both of which have $g_{a\gamma\gamma} \sim \alpha/(2\pi f_a) \sim \alpha m_a/(2\pi \Lambda_\text{QCD}^2)$.
}
  \label{fig:gaGG}
\end{figure}

A coupling of axions to photons, $\mathcal{L} \supset g_{a\gamma\gamma} aF \tilde F$,  is also generated due to the electromagnetic anomaly when rotating the phases of the first two generations of quarks. Comparing with Eq.~\ref{couplings1},
\begin{align}
g_{a\gamma\gamma} = \frac{\alpha}{2\pi} \frac{1}{f} \,.
\end{align}
The experimental constraints on this coupling are shown in Figure \ref{fig:gaGG}. 
We see that this construction generically relates the axion-nucleon coupling to the axion-photon coupling and therefore one would expect positive signals to appear in both classes of experiments.\footnote{
Note, however, that one can also construct models in which the photon couplings are absent by charging additional fields under \pq in such a way to avoid the EM anomaly for \pq. In this case, only the experiments sensitive to nuclear couplings would see a signal.
}

\section{Enhancing the hadronic couplings of the QCD axion}
\label{section:clockwork}
The axion model in the previous section clearly does not solve the strong CP problem, since the axion does not couple to gluons. In this section we explore how one can extend the mechanism used in the previous section to obtain a QCD axion with axion-nucleon couplings parametrically larger than its coupling to gluons. This can be achieved by including extra colored fermions with chiral charges under \pq to generate the $a G \tilde G$ coupling. However, in order to get parametric separation between $g_{aNN}$ and $m_a$, the \pq charge of SM fermions has to be parametrically larger than the charge of the new colored fermion. One way to achieve this in a controlled way is through the clockwork mechanism~\cite{Choi:2015fiu,Kaplan:2015fuy}. A similar construction has already been used to parametrically enhance the axion coupling to photons in Ref.~\cite{Farina:2016tgd}~(see also Ref.~\cite{Agrawal:2017cmd} for extensions of this idea).

The clockwork mechanism introduces $N+1$ complex scalars, with a potential
\begin{align}
V(\phi_0, \ldots, \phi_N) = \sum_{j=0}^N \left(-m^2 |\phi_j|^2 + \lambda|\phi_j|^4 \right) + \sum_{j=0}^{N-1} \left(\kappa \phi_j^\dagger \phi_{j+1}^3 + h.c. \right) \, ,
\label{eq:clockwork-potential}
\end{align}
with $\kappa \ll \lambda$. In the limit $\kappa \rightarrow 0$ the potential respects a $U(1)^{N+1}$ global symmetry, which is spontaneously broken by
each scalar getting a VEV $f = m \sqrt{2/\lambda}$. This leads to $N+1$ goldstone bosons $\pi_i$ parametrized by $\phi_j \rightarrow \frac{f}{\sqrt{2}} e^{-\frac{i\pi_j}{f}}$. The $\kappa$ term in Eq.~\ref{eq:clockwork-potential} explicitly breaks $U(1)^{N+1}$ down to a single $U(1)$, generating masses $m_j \sim \kappa f^2/2$ for $N$ of the would-be goldstone modes. One mode, 
\begin{align}
a  \propto \sum_{j=0}^N \frac{1}{3^j} \pi_j \,,
\label{eq:clockwork-localization}
\end{align}
remains massless, corresponding to the remaining $U(1)$ symmetry that is not broken by the $\kappa$ terms.

The unbroken $U(1)$ is identified with $U(1)_{PQ}$, but the approximate $U(1)_0 \times U(1)_1 \times \dots U(1)_{N}$ symmetry structure is very useful to write down models in which fields have exponentially different \pq charges. In particular, one can show that a field with unit charge under $U(1)_k$ has a \pq charge $3^k$ smaller than a field with unit charge under $U(1)_0$~(see e.g. Ref.~\cite{Kaplan:2015fuy}). In order to obtain large quark derivative couplings we take the SM quark singlets to be charged under $U(1)_0$, and the new colored fermions to be charged under $U(1)_k$, with $0<k\leq N$. The Yukawa sector of the model can be written as
\begin{align}
\mathcal{L} \supset \mathcal{L}_{qH} |_{\phi\rightarrow\phi_0} + y_\psi \phi_k \psi \psi^c + h.c.
\end{align}
where $\psi$ and $\psi^c$ are non-SM fermions that are charged under QCD, and $\mathcal{L}_{qH}|_{\phi\rightarrow\phi_0}$ is the Lagrangian in Eq.~\ref{eq:lagrangian-symmetry} with the original $\phi$ field replaced by $\phi_0$. 

In this setup, the new fermions get a mass $\sim y_\psi f$, and can be integrated out. More importantly, due to the $\psi$ coupling to $\phi_k$, integrating out these fermions generates an axion coupling to gluons with
\begin{align}
\frac{a}{3^k f} \frac{\alpha_s}{8\pi} G \tilde{G} \,,
\end{align}
where the $3^k$ factor can be understood from Eq.~\ref{eq:clockwork-localization} (or directly from the fact that $\psi$ has \pq charge $3^{-k}$). As a consequence, at energies below $\Lambda_{QCD}$ the axion gets a potential $V \sim m_\pi^2 f_\pi^2 \cos(\frac{a}{3^k f})$ and therefore a mass $m_a \sim \frac{m_\pi f_\pi}{3^k f}$. Its couplings to photons, and its derivative couplings to quarks and nucleons follow from the previous section. These couplings are all set by the scale $\frac{1}{f}$ and hence can be parametrically larger than $\frac{1}{3^k f} = \frac{1}{f_a}$ which is the generic expectation for a given axion mass. Bounds on the axion-nucleon coupling specific to the QCD axion are similar to those for non-QCD axions (as shown in Figure \ref{fig:gaNN}), with a few exceptions we discuss below. 

Firstly, the bounds from black hole superradiance \cite{Arvanitaki:2014wva} over the range of masses probed are expected to change. The upper limit on the superradiance bounds in Figure~\ref{fig:gaNN} comes from the fact that for non-QCD axions, larger $g_{aNN}$ implies larger axion self-interactions (assuming the potential is similar to Eq.~\ref{eq:alp-potential}), which hinders the growth of the boson cloud. This is no longer true for the QCD axion model described in this section, where the full axion potential is essentially independent of $g_{aNN}$. In this case the main effects that could inhibit the axion cloud growth are interactions with matter in the accretion disk or with magnetic fields surrounding the black hole. Calculating the impact of these interactions is beyond the scope of this work, but very simple estimates indicate that the superradiance bounds should extend to larger $g_{aNN}$ for QCD axions compared to those shown in Figure~\ref{fig:gaNN}.

\begin{figure}[t]
  \centering
    \includegraphics[height=5.3cm]{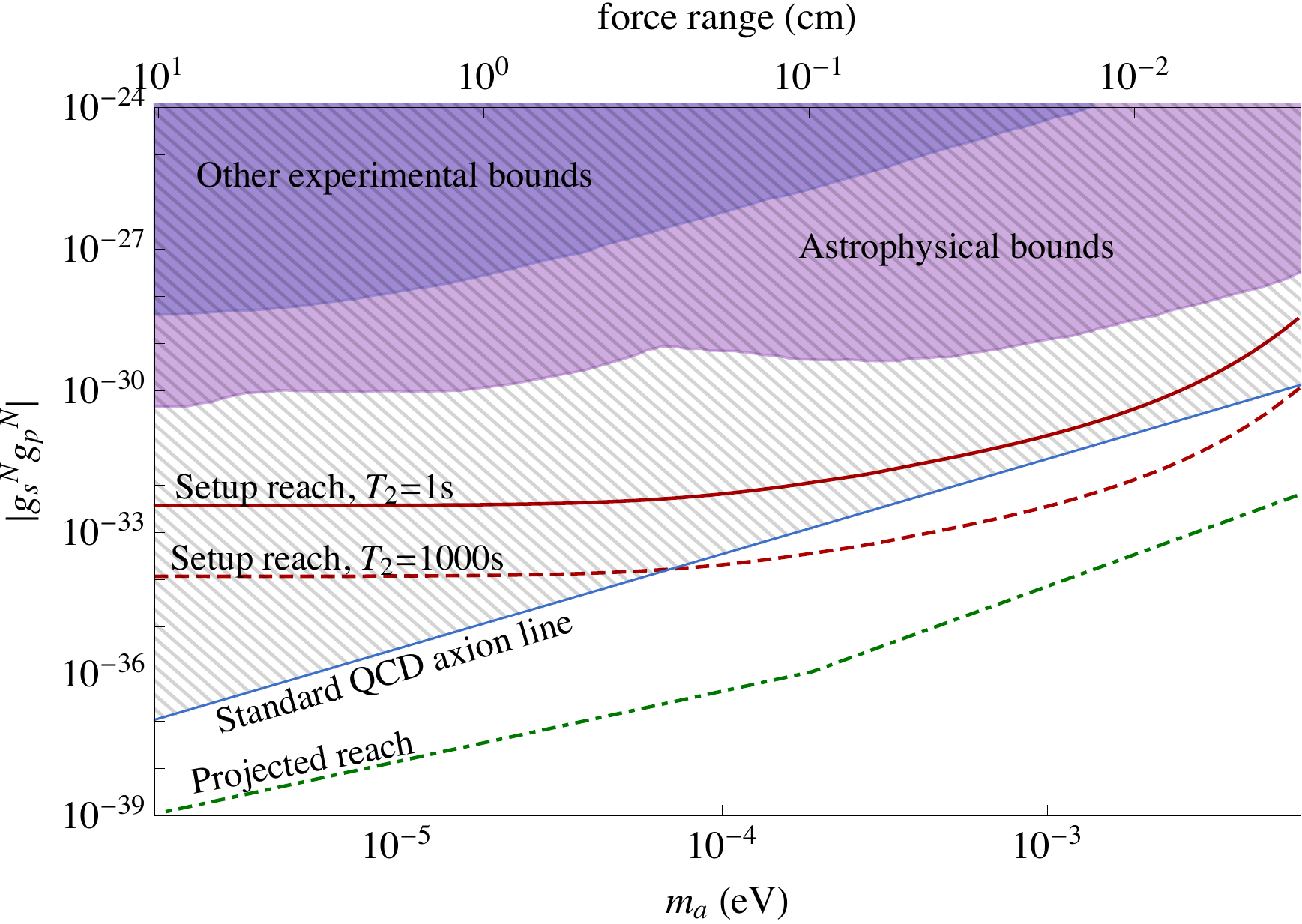}
     \includegraphics[height=5.3cm]{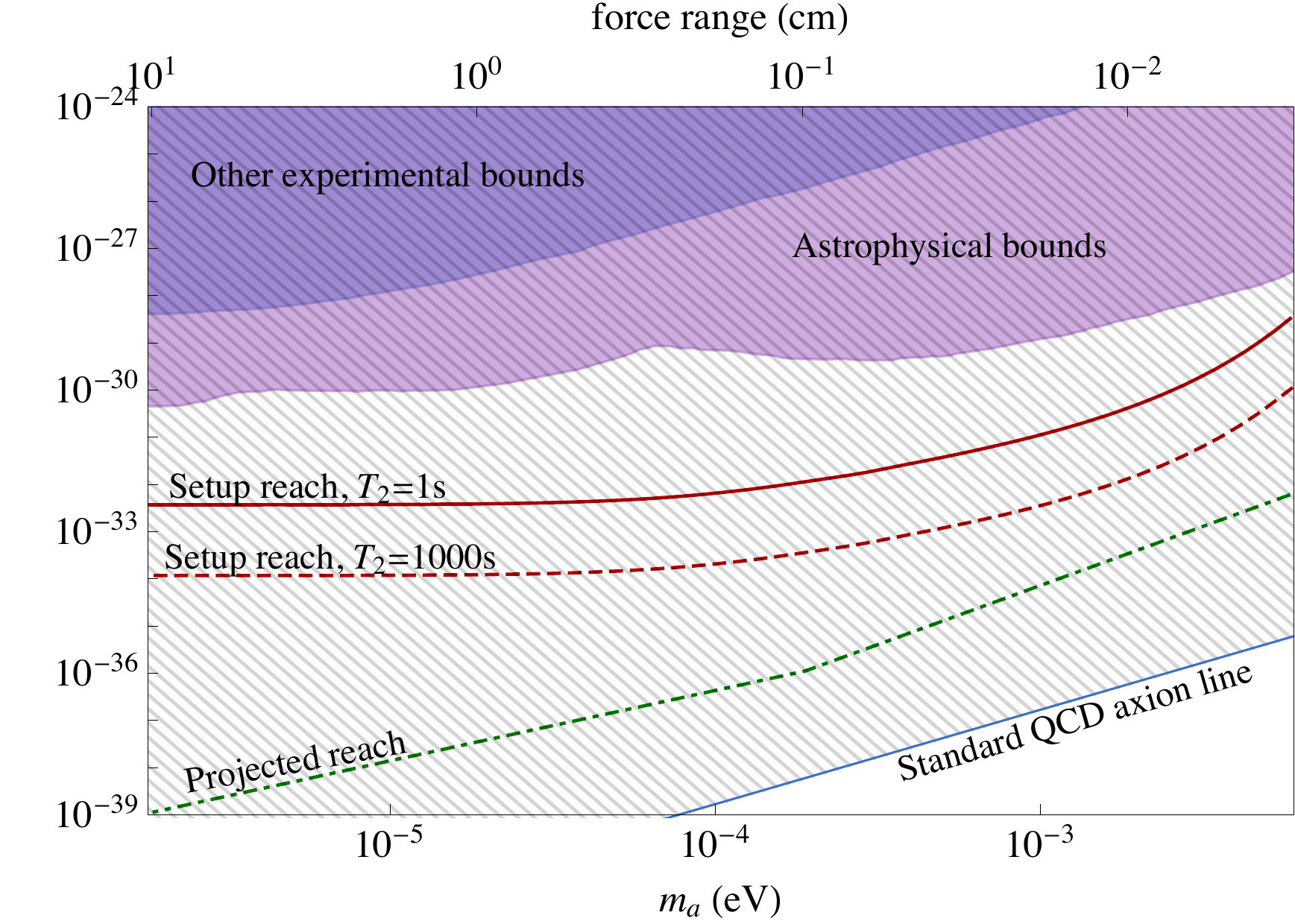}
   \caption{Bounds on the monopole-dipole coupling between nucleons from QCD axion-mediated forces. The solid and dashed red curves are the projected sensitivities of the ARIADNE setup for various settings, discussed in Ref. \cite{Arvanitaki:2014dfa}. The green dot-dashed curve is a future projection for a scaled-up setup. Other bounds \cite{PhysRevD.86.015001} come from combined $g_s^N$ and $g_p^N$ limits. The solid blue line is the monopole-dipole coupling for a standard QCD axion assuming $|g_s|\sim 10^{-21} (m_a/10^{-4} \text{ eV})$ (left) and $|g_s|\sim 10^{-26} (m_a/10^{-4} \text{ eV})$ (right). The parameter space for our QCD axion model (hashed region) extends above the standard QCD axion line.
   }
  \label{fig:ariadne}
\end{figure}

Secondly, the ARIADNE experiment \cite{Arvanitaki:2014dfa} will be sensitive to a large portion of unconstrained parameter space and is not shown in Figure \ref{fig:gaNN}. This experiment is sensitive to axion mediated forces between nucleons \cite{PhysRevD.30.130}, and thus its sensitivity is independent of whether axions form a sizable fraction of dark matter. The experiment will be sensitive to both monopole-dipole and dipole-dipole interactions on length scales of $1/m_a$. The monopole interaction requires a non-zero CP-violating $\theta G\tilde{G}$ term (i.e. only possible with a QCD axion), since it does not arise from a shift-symmetric coupling. It also requires that the minimum of the axion potential is not exactly at $\theta=0$, since the monopole coupling, $g_s^N$, is proportional to the $\theta$ angle.
The dipole coupling, $g_p^N$, is model-dependent and is directly related to the axion-nucleon derivative coupling, $g_{aNN}$. In our model, the dipole coupling can be parametrically larger than that for standard axions ($g_p^N \sim m_N/f_a$). The bounds on monopole-dipole couplings $|g_s^N g_p^N|$ are shown in Figure \ref{fig:ariadne}, while the sensitivity to the dipole-dipole interactions is expected to be in a region of parameter space already in tension with astrophysical bounds. We again see that the parameter space for our axion model (hashed region) extends significantly above the standard QCD axion line, already within reach of the proposed ARIADNE setup for a wide range of values for $\theta$.

\section{Conclusion}

In this work we explore models where the axion coupling to nucleons can be much larger than what is generally expected for a given axion mass. This motivates searching for axions in a much wider range of parameter space than what typical axion models suggest. The mechanism to enhance the axion-nucleon coupling has a number of interesting properties. It predicts that axion couplings are not flavor diagonal and therefore can induce flavor changing neutral currents. Some of these processes lead to constraints that are comparable to astrophysical bounds (which contain large uncertainties). It also predicts that the axion couples with opposite sign to protons and neutrons, which could potentially lead to interesting features that were not explored in this work.

We also explicitly construct a QCD axion model in which the axion-nucleon derivative coupling is parametrically larger than the axion coupling to gluons that is associated with solving the strong CP problem. This provides a concrete example in which the derivative coupling to nucleons is essentially independent from the decay constant associated with the axion potential. Models of this kind significantly expand the QCD axion parameter space that can be tested by experiments sensitive to the axion-nucleon coupling.

\section*{Acknowledgments}
We thank Anson Hook, Asimina Arvanitaki, Diego Redigolo, Gilad Perez, Martin Schmaltz, Masha Baryakhtar, Peter Graham, Prateek Agrawal, Savas Dimopoulos, and Zoltan Ligeti for useful conversations. We thank Sebastian Ellis for useful discussions and comments on the draft. G.M.T. acknowledges the support of NSF grant No. PHY-1720397 and DOE Early Career Award DE-SC0012012. M.T. is supported by the Stanford Graduate Fellowship.

\bibliographystyle{utphys}
\bibliography{ref}
\end{document}